\documentclass[12pt]{iopart}
\usepackage[nolist,nohyperlinks]{acronym}
\usepackage{amsbsy}
\usepackage{latexsym}
\usepackage{graphicx}
\usepackage{url}
\usepackage{xcolor}
\usepackage{bm}
\usepackage{caption}

\newcommand{\D}{\mathrm{d}}

\acrodef{GW}{gravitational wave}
\acrodef{BH}{black hole}
\acrodef{BBH}{binary black holes}
\acrodef{NS}{neutron star}
\acrodef{BNS}{binary neutron stars}
\acrodef{NSBH}{neutron star--black hole}
\acrodef{IMBH}{intermediate mass binary black holes}
\acrodef{NS}{neutron star}
\acrodef{MC}{Monte-Carlo}
\acrodef{SNR}{signal-to-noise ratio}

\begin{document}

\title[Estimation of the Sensitive Volume for Gravitational-wave Source Populations]{Estimation of the Sensitive Volume for Gravitational-wave Source Populations Using Weighted Monte Carlo Integration}

\author{Vaibhav Tiwari}
\address{Cardiff School of Physics and Astronomy,
Cardiff University, Queens Buildings, The Parade, Cardiff CF24 3AA, UK}
\vspace{10pt}
\ead{tiwariv@cardiff.ac.uk}
\begin{indented}
\item[]
\end{indented}

\begin{abstract}
The population analysis and estimation of merger rates of compact binaries is one of the important topics in gravitational wave astronomy. The primary ingredient in these analyses is the population-averaged sensitive 
volume. Typically, sensitive volume, of a given search to a given simulated source population, is estimated by drawing signals from the population model and adding them to the detector data as injections. Subsequently injections, 
which are simulated gravitational waveforms, are searched for by the search pipelines and their \ac{SNR} is determined. Sensitive volume is estimated, by using \ac{MC} integration, from the total number of injections added to the data, 
the number of injections that cross a chosen threshold on \ac{SNR} and the astrophysical volume in which the injections are placed. So far, only fixed population models have been used in the estimation of \ac{BBH} merger rates. However, 
as the scope of population analysis broaden in terms of the methodologies and source properties considered, due to an increase in the number of observed \ac{GW} signals, the procedure will need to be repeated multiple times at a large 
computational cost. In this letter we address the problem by performing a weighted \ac{MC} integration. We show how a single set of generic injections can be weighted to estimate the sensitive volume for multiple population models; 
thereby greatly reducing the computational cost. The weights in this \ac{MC} integral are the ratios of the output probabilities, determined by the population model and standard cosmology, and the injection probability, determined 
by the distribution function of the generic injections. Unlike analytical/semi-analytical methods, which usually estimate sensitive volume using single detector sensitivity, the method is accurate within statistical errors, comes at 
no added cost and requires minimal computational resources. 
\end{abstract}

%
\vspace{2pc}
\section{Introduction}

Understanding source population or a mixture of them is one of the primary goal of gravitational wave astronomy. These studies include population analysis, estimating rate of occurrence of an astrophysical phenomenon or placing 
upper-limit on the occurrence of a proposed astrophysical phenomenon in the event of a null observation. Prior to the  observation of GW150914 \cite{gw150914} LIGO published upper limits on the merger rate of stellar 
mass compact binaries after all the scientific runs. These include upper limits, as a function of total mass for \ac{BBH} and \ac{NSBH} binaries \cite{s3bbh}, for binaries with first mass fixed at 1.35M$_\odot$ and second 
mass uniformly distributed between 2M$_\odot$ and 20M$_\odot$ \cite{s3s4bbh}, for fixed mass \ac{BNS}, \ac{BBH} and \ac{NSBH} binaries \cite{s5bbh}, and for \ac{BBH} on the component mass plane \cite{s6bbh, s6lmbbh}. 
The observation of GW150914 during LIGO's first observation run and the observation of GW170817 during the second observation run provided opportunity for the first time for the estimation of the \ac{BBH} and the \ac{BNS} merger rate 
\cite{Abbott:2016nhf, first_bns}. For example, the rate upper limit on \ac{BNS} mergers, estimated at the end of LIGO's sixth scientific run, was $\sim 10^{-4} \mathrm{Mpc}^{-3} \mathrm{yr}^{-1}$ at 90\% confidence, while the 
\ac{BNS} merger rate estimated after the observation of GW170817 is $\sim 10^{-6} \mathrm{Mpc}^{-3} \mathrm{yr}^{-1}$ \cite{s6lmbbh, first_bns}.

Sources such as \ac{IMBH} and \ac{NSBH} are anticipated for observation and as of yet only upper limits on the merger rate have been published \cite{s5imbh, s6imbh}. Estimated rate limit, for 
\ac{IMBH} sources, at the end of LIGO's first observation run, is $0.93 \mathrm{Gpc}^{-3} \mathrm{yr}^{-1}$, while for \ac{NSBH} binaries it is $3600 \mathrm{Gpc}^{-3} \mathrm{yr}^{-1}$ \cite{O1IMBH, postO1IMBH, o1nsbh}. Rate limit on eccentric 
binaries have also been estimated \cite{s6ebbh}.

The primary ingredient in the estimation of merger rate or upper-limit on the merger rate \cite{rlimit} is the sensitive volume. In the event of a non-detection, upper rate limits have been placed for source populations. Under the assumption that 
observation of a \ac{GW} signal is a Poisson process, the rate limit, quoted for a chosen confidence interval, is inversely proportional to the sensitive volume of the source population. For a source population of compact binaries, which has 
resulted in an observation, if the intrinsic, redshift independent, rate of coalescence is $R$, then one can expect to observe $R\langle V T\rangle$ number of \ac{GW} signals. But, as  the intrinsic rate is not known the variables are reversed to 
estimate the expected rate \cite{Abbott:2016nhf, lsc_binary_2016}, 
\begin{equation} R = \frac{N_{\mathrm{obs}}}{\langle V T\rangle}, \end{equation}
where $\langle V T\rangle$ is the time-volume product, computed using the population averaged sensitive volume $\langle V \rangle$ and time $T$ during which $N_{\mathrm{obs}}$ observations have been made. One can perform a more 
sophisticated rate estimate e.g.\ while including the confidence with which observations have been made \cite{Farr:2015}, but the primary ingredient is the sensitive volume. 

Observation of the first \ac{GW} signal also opened the window for population analysis. The first result estimated the probability distribution of the exponent under the assumption that mass distribution of the black holes 
follows a power-law distribution \cite{lsc_binary_2016}. Since then additional work has gone into understanding the spin distribution of the black holes \cite{spin1, spin2}. With the increase in the number of observations, 
it is expected, that the scope of population analysis will broaden in terms of methodologies and source properties considered. Sensitive volume is a primary ingredient in the Bayesian framework of population analysis as 
it accounts for the selection effects \cite{sel_bias, lsc_binary_2016}. A range of sensitive volumes, corresponding to different population models, are required. Large scale simulations for the estimation of sensitive volume is computationally 
expensive and performing multiple analyses dedicated at the source populations is not computationally viable.

\section{Estimation of sensitive volume of a search}
The population averaged time-volume is defined as,
\begin{equation}
 \langle V T\rangle = \int \D z \D\theta \;\frac{\D V_c}{\D z} \frac{1}{1+z} p_{\mathrm{pop}}(\theta)f(z, \theta)\cdot T, \label{eq:vis_vol}
\end{equation}
where $\D V_c/\D z$ is the differential co-moving volume, $p_{\mathrm{pop}}(\theta)$ is the distribution function for the astrophysical population and $f(z, \theta)$ is the probability of recovering a signal, with parameters $\theta$, at 
a redshift $z$. The additional factor of $1 + z$ in the denominator accounts for the time-dilation in the intrinsic rate caused by the expanding universe \cite{Abbott:2016nhf}. Equation \ref{eq:vis_vol} is estimated by using Monte-Carlo 
integration \cite{MCint}. Signals are drawn from the population model and added to the detector data as injections. The injections are placed in redshift as determined by the standard cosmology. The probability of making an injection at 
redshift $z$ is given by,
\begin{equation}
 p(z) = \frac{\D V_c/\D z}{(1+z) V_0}, 
 \label{preds}
\end{equation}
where $V_0$ is the astrophysical volume in which injections have been made corresponding to a maximum redshift $z_{\mathrm{max}}$,
\begin{equation}
 V_0 = \int_0^{z_\mathrm{max}} \frac{\D V_c}{\D z} \frac{1}{1+z} \;\D z. \label{eq:V0}
\end{equation}
The injections are searched for by search pipelines and their \ac{SNR} is determined \cite{pycbc, titopycbc, gstlal, cwb}. An injection is categorized as recovered if it crosses a chosen \ac{SNR} threshold. Sensitive volume, in terms of the 
number of injected signals $N_{\mathrm{inj}}$ and the number of recovered signals $N_{\mathrm{rec}}$, is given by \cite{Abbott:2016nhf,lsc_binary_2016},
\begin{equation}
 \langle V \rangle = V_0 \frac{N_{\mathrm{rec}}}{N_{\mathrm{inj}}}. \label{eq:vol_mc}
\end{equation}

Such an analysis suffers from two drawbacks: (i) The distribution function $s(\theta)$ is fixed, hence, multiple injection runs will be required if sensitive volume is needed for a range of population models, and (ii) as 
the redshift placement of the injections is independent of the parameters $\theta$, significant number of injections don't contribute, due to being placed at redshift values where their recovery is impossible. 

In the following section we overcome these limitations by performing a weighted \ac{MC} integration on a set of generic injections. In Section \ref{sec:analysis} we describe the analysis and present results 
and conclusion in Section \ref{sec:results}.

\section{Analysis}
\label{sec:analysis}

The shortcomings listed in the previous section can be overcome by using generic injection sets followed by weighting the injections in the \ac{MC} integration. The applied weight is the 
ratio of the output probability, determined by the population model and standard cosmology, and the injection probability, determined by the distribution function of the generic injections. This allows the use of the same set of injections in the 
estimation of sensitive volume for different population models. The weight $w$ on the $i^{th}$ injection reads,
\begin{equation}
  w_i = \frac{p_{\mathrm{pop}}(\{m_1, m_2, \vec{s}_1, \vec{s}_2\}_i)\cdot p\left(z(d_i)\right)}{p_{\mathrm{inj}}\left(\{m^d_1, m^d_2, \vec{s}_1, \vec{s}_2, d\}_i \right)\cdot J_i}.
  \label{weights}
\end{equation}
The output probability, in the numerator, depends on the probability distribution of the target population model $p_{\mathrm{pop}}$, defined in terms of the source frame masses ($m_1$, $m_2$) and the dimensionless spins 
($\vec{s}_{1, 2} \equiv \{s_x, s_y, s_z\}_{1, 2}$), and the redshift probability, determined by the standard cosmology and as defined in Equation \ref{preds}. The maximum redshift, $z_\mathrm{max}$, defines the boundary beyond which events produced by the population model are not recoverable by the detector network. The input probability $p_{\mathrm{inj}}$, in the denominator, depends on the injection 
distribution and is defined in terms of the detector frame masses ($m^d_1$, $m^d_2$), spins ($\vec{s}_{1, 2}$) and luminosity distance $d$. The Jacobian, $J$, maps the probability from the detector frame parameters, ($m_1^d$, $m_2^d$, $d$), to the 
source frame parameters ($m_1$, $m_2$, $z$),
\begin{equation}
J_i = (1 + z(d_i))^2 \left(\frac{\partial d}{\partial z}\right)_i.
\end{equation}
Due to red-shifting, the signal produced from a binary with masses $m_1$ and $m_2$ is observed in the detector frame with masses $m^d_1 = (1 + z)\;m_1$ and 
$m^d_2 = (1 + z)\;m_2$. Moreover, if a binary merges at a redshift $z$, the merger is observed, by the detector, at the corresponding luminosity distance. The amplitude of the signal, as observed by the detector, is inversely proportional to the 
luminosity distance. Hence, it is convenient to perform injections using detector 
based quantities. 

Injections should adequately cover the parameter space i.e. sufficient number of injections should be made, in parts of the parameter space, where $p_{\mathrm{pop}}$ is non-negligible. Moreover, redshift placement should ensure that injections are 
placed wherever the recovery probability, $f(z, \theta)$, is non-zero. It is better to have a quantitative 
measure reflecting the goodness of coverage, but, however, for the population models that are only dependent on the component masses a good assessment can be based on how well injections cover the population model on the source frame component mass-plane. Adequate coverage in redshift can be ascertained by injecting distance uniform in chirp distance with a large enough fiducial distance.

This analysis does not need dedicated injection runs. Injection runs are regularly performed to assess efficiency of the search pipelines in different regions of the parameter space \cite{pycbc,titopycbc, gstlal, cwb}. Table 
\ref{table:inj_strat} summarizes the most commonly used parameter distributions used in making injections. Usually, multiple injection sets, covering different regions of masses and spins, are used to adequately cover the parameter space and 
are ideal for the use of the presented analysis (if the same number of injections are performed from $k$ injections set, the input probability is summed over all the injection distributions in the set, i.e. 
$p_{\mathrm{inj}}(m_1^d, m_2^d, s_{1z}, s_{2z}, d) = \sum_{j = 1}^k p^j_{\mathrm{inj}}(m_1^d, m_2^d, s_{1z}, s_{2z}, d)$, $p^j_{\mathrm{inj}}(\cdots) = p^j(m_1^d, m_2^d)\cdot p^j(s_{1z})\cdot p^j(s_{2z}) \cdot p^j(d)$ being the injection 
distributions for the $j^{th}$ injection set).
\begin{table}[ht]
\centering 
\begin{tabular}{cc}
\hline
 Distribution uniform in& Probability Density\\
 \hline
 Component Mass & $p(m_1^d, m_2^d) = \frac{1}{\left(m^d_{\mathrm{max}} - m^d_{\mathrm{min}}\right)^2}$ \\[7pt]
 Total Mass & $p(m_1^d, m_2^d)  = \frac{1}{(m^d_{\mathrm{max}} - m^d_{\mathrm{min}}) (m^d_1 + m^d_2 - 2 m^d_{\mathrm{min}}) }$ \\[7pt]
 Aligned Spin & $ p(s_z) = \frac{1}{s_{z,\mathrm{max}} - s_{z,\mathrm{min}}}$ \\[7pt]
 Total Spin & $ p(s_z) = \frac{\log(s_{\mathrm{max}} - \log(|s_z|))}{2 s_{\mathrm{max}}}$ \\[7pt]
 Distance & $p(d) = \frac{1}{d_{\mathrm{max}} - d_{\mathrm{min}}}$ \\[7pt]
 Chirp Distance & $p(d) = \frac{\mathcal{M}_{\mathrm{BNS}}^{(5/6)}}{\left(\mathcal{M}^d\right)^{(5/6)}(d_{\mathrm{max}} - d_{\mathrm{min}})}$ \\[7pt]
\hline
\end{tabular}
\caption{The probability density of some of the mass, spin and distance distribution used in performing injections. Typically an injection run will select a distribution for mass, spin and distance, each. For the first distribution, in this table, 
both the component mases are uniformly distributed between the minimum value, $m^d_{\mathrm{min}}$, and the maximum value, $m^d_{\mathrm{max}}$. For the second distribution, both the component masses have the same minimum and the maximum value. 
$m_1^d$ is uniformly distributed between $m^d_{\mathrm{min}}$ and $M^d - m^d_{\mathrm{min}}$. The total mass, $M^d =m_1^d + m_2^d$, is uniformly distributed between $M^d_{\mathrm{min}}=2m^d_{\mathrm{min}}$ and 
$M^d_{\mathrm{max}} = m^d_{\mathrm{min}} + m^d_{\mathrm{max}}$. For the third distribution, pertaining to spins, $s_z$ is uniformly distributed between $s_{z,\mathrm{min}}$ and $s_{z,\mathrm{max}}$; other spin components 
are zero. For the fourth distribution, spin magnitudes are uniformly distributed between $s_{\mathrm{min}}$ and $s_{\mathrm{max}}$; other spin components follow the same distribution with $s_z$ replaced by $s_x$ or $s_y$. 
For the fifth distribution, pertaining to luminosity distance, $d$ is uniformly distributed between $d_{\mathrm{min}}$ and $d_{\mathrm{max}}$. For the sixth distribution, $d$ is based on the mass distribution and is 
uniformly distributed between $d_{\mathrm{min}}\left(\mathcal{M}^d\right)^{(5/6)}$ and $d_{\mathrm{max}}\left(\mathcal{M}^d\right)^{(5/6)}$, where $d_{\mathrm{min}}$ and $d_{\mathrm{min}}$ are fixed minimum and maximum 
fiducial distances, (called as chirp distance in technical publications). $\mathcal{M}_{\mathrm{BNS}}$ is the chirp mass of a 1.4M$_\odot$ - 1.4M$_\odot$ \ac{BNS}, where chirp mass is defined as, 
$\mathcal{M}^d = (m^d_1m^d_2)^{(3/5)}/(m^d_1 + m^d_2)^{(1/5)}$. This distribution exploits the approximate dependence of the detectability of a binary on its chirp mass to avoid placement of significant number of injections at redshifts where 
their recovery is impossible (the maximum distance at which an injection is recovered at a high enough \ac{SNR} to be detectable is roughly dependent on the chirp mass of the binary).}
\label{table:inj_strat}
\end{table}

Weighting maps the injections to the population model and cosmological redshift distribution, but, astrophysical volume in which injections are made can not be estimated using Equation \ref{eq:V0}. Input probability 
can also be expressed in terms of the parameters $(\mathcal{M}^d, q, \vec{s_1}, \vec{s}_s)$, where $q = m_2/m_1$ is the mass ratio. Out of these parameters, only parameter that changes value from source to detector frame is the chirp mass. Unlike 
the injections sampled from the population model, that cut a rectangular shape, the generic injections are characterised by the detector frame quantities and cut an irregular shape in the $\mathcal{M} - z$ plane. Redshift 
placement of the injections depend on the source frame chirp mass. The astrophysical volume in which injections are made is given by,
\begin{equation}
V_0^m = \int_{\mathcal{M}_{\mathrm{min}}}^{\mathcal{M}_{\mathrm{max}}}\int_{z_{\mathrm{min}}(\mathcal{M})}^{z_{\mathrm{max}}(\mathcal{M})} p_{\mathrm{pop}}(\mathcal{M}) \frac{\D V_c}{\D z} \frac{1}{1+z}\D z \; 
        \D \mathcal{M},
\label{eq:mod_V0}
\end{equation}
where $p_{\mathrm{pop}}(\mathcal{M})$ is the population model expressed as a function of source frame chirp mass after being marginalized over other parameters. The boundaries $z_{\mathrm{max}}(\mathcal{M})$ and 
$z_{\mathrm{min}}(\mathcal{M})$ are determined by the injection sets, while the boundaries $\mathcal{M}_{\mathrm{min}}$ and $\mathcal{M}_{\mathrm{max}}$ are set by the population model. Estimation of Equation $V_0^m$ is 
not straightforward and we resort to \ac{MC} integration again. We simulate a physical source population by sampling chirp masses from the population model and redshifts according to distribution in Equation \ref{preds}. We calculate the 
detector frame chirp mass and luminosity distances of these samples and count the number of samples that have non-zero injection probability in the $\mathcal{M}^d-d$ space ($p_{\mathrm{inj}}$ expressed as a function of 
$\mathcal{M}^d$ and $d$ and marginalized over other parameters). The \ac{MC} estimate of $V_0^m$ is given by,
\begin{equation}
 V_0^m = V_0 \frac{N_{\mathrm{inside}}}{N_\mathrm{samples}},
 \label{v0m}
\end{equation}
where $N_{\mathrm{inside}}$ is the total number of samples that have non-zero injection probability and $N_\mathrm{samples}$ is the total number of samples drawn from the population model. $V_0$ is given by Equation 
\ref{eq:V0} corresponding to the maximum redshift $z_\mathrm{max}$. If the computational cost of sampling chirp mass is low, one can generate 
a large number of samples to safely ignore statistical errors incurred in this \ac{MC} integration.
Finally, sensitive volume is estimated by putting Equation \ref{weights} and Equation \ref{v0m} together,
\begin{equation}
 \langle V \rangle = V_0^m\; \frac{\displaystyle \sum_{i\;\in\;\mathrm{Rec}}\;w_i}{\displaystyle \sum_{i\; \in \;\mathrm{Inj}} w_i}. \label{eq:vol_mc_m}
\end{equation}

The \ac{MC} integration has errors associated with it. The error, in the mean weight, in the denominator of Equation \ref{eq:vol_mc_m} is given by,
\begin{equation}
 \left(N_{\mathrm{inj}}\delta w\right)^2 = \frac{\displaystyle N_{\;\mathrm{inj}}\sum_{i\;\in\;\mathrm{inj}} w_i^2 - \left( \displaystyle \sum_{i\;\in\; \mathrm{inj}} w_i\right)^2}{N_{\mathrm{inj}}}.
\end{equation}
The same expression holds true for the numerator but summed over recovered events (the weight for a missed injection is zero). Propagating the errors to find the error in the ratio gives
\begin{equation}
\frac{\delta \langle V \rangle}{\langle V \rangle} = \sqrt{\displaystyle \sum_{k = \mathrm{inj}, \mathrm{rec}} \frac{\displaystyle N_{\mathrm{inj}}\sum_{i\;\in\;k} w_i^2 - 
                                                                                           \left( \displaystyle \sum_{i\;\in\;k} w_i\right)^2}{N_{\mathrm{inj}}\left(\displaystyle \sum_{i\;\in\;k} w_i\right)^2}.}
\label{eq:bi_err}
\end{equation}
$N_{\mathrm{inj}}$ is of the order of few thousands and the number of the recovered events is only a small fraction of the injected events, additionally, irrespective of the value of the weights, the sum of squares of the weights is much less than 
the square of the sum of the weights. Under these conditions the dominant term is $\sum_{\mathrm{Rec}} w_i^2/(\sum_{\mathrm{Rec}} w_i)^2$ (for example, with a total 100000 injections and ten percent recovered injections, this term is around ten 
times the remaining terms). The error in the sensitive volume reduces to,
\begin{equation}
 \frac{\delta \langle V \rangle}{\langle V \rangle} = \frac{\sqrt{\displaystyle \sum_{\mathrm{Rec}} w_i^2}}{\displaystyle \sum_{\mathrm{Rec}} w_i}.
\end{equation}
In Section \ref{sec:results} we apply the algorithm and obtain some results.
\section{Results and Conclusion}
\label{sec:results}
In this section we estimate the time-volume product for the same stretch of data that was used in estimating an updated \ac{BBH} merger rate after the \ac{GW} observation GW170104 \cite{GW170104}. The results discussed in this section are based on 
the injection runs performed using the search pipeline PyCBC \cite{pycbc,titopycbc}.

So far the two population models that have been used in the estimation of \ac{BBH} merger rate are as follows \cite{lsc_binary_2016}:
\begin{enumerate}
\item[a] Uniform in the logarithm of the component masses, with combined probability density given by
$p(m_{1}, m_{2}) \propto {m_{1}}^{-1}{m_{2}}^{-1}$ and
\item[b] The primary mass follows a power-law distribution while the secondary mass is uniformly distributed between smallest mass and the primary mass,
$p(m_{1}) \propto m_{1}^{-2.35}$ with a uniform distribution
on the second mass. 
\end{enumerate}
The masses are required to be $5 \; \mathrm{M}_\odot \leq m_2 \leq m_1$ and $m_1 + m_2 \leq 100 \; \mathrm{M}_\odot$. Spins have been chosen to be aligned and uniformly distributed between -0.99 and 0.99.
Sensitive volume has been estimated by performing \ac{MC} integral using injections with parameters directly sampled from the population models. The statistical error in the MC integration for this analysis is 
$\delta \langle V \rangle = \sqrt{\epsilon(1-\epsilon)/N_{\mathrm{Inj}}}$, where $\epsilon$ is the efficiency of recovering injections.

We also estimate the sensitive volume, using the weighted MC integration. We use six different injection distributions. For all the distributions, the injections probability of both the component masses is uniform over the logarithm of the component 
masses ($p(m_1) \propto m^{-1}$), with component mass range: 
\begin{enumerate}
\item[a] Component mass range: 4M$_\odot$ to 180M$_\odot$,
\item[b] Component mass range: 6M$_\odot$ to 180M$_\odot$,
\item[c] Component mass range: 10M$_\odot$ to 180M$_\odot$,
\item[d] Component mass range: 16M$_\odot$ to 180M$_\odot$, 
\item[e] Component mass range: 24M$_\odot$ to 180M$_\odot$,
\item[f] Component mass range: 36M$_\odot$ to 180M$_\odot$.
\end{enumerate}
Spin distribution is aligned and uniformly distributed between -0.99 and 0.99, hence, the weights in the \ac{MC} integral depends only on the component masses.

\begin{figure}
\includegraphics[width=0.95\textwidth]{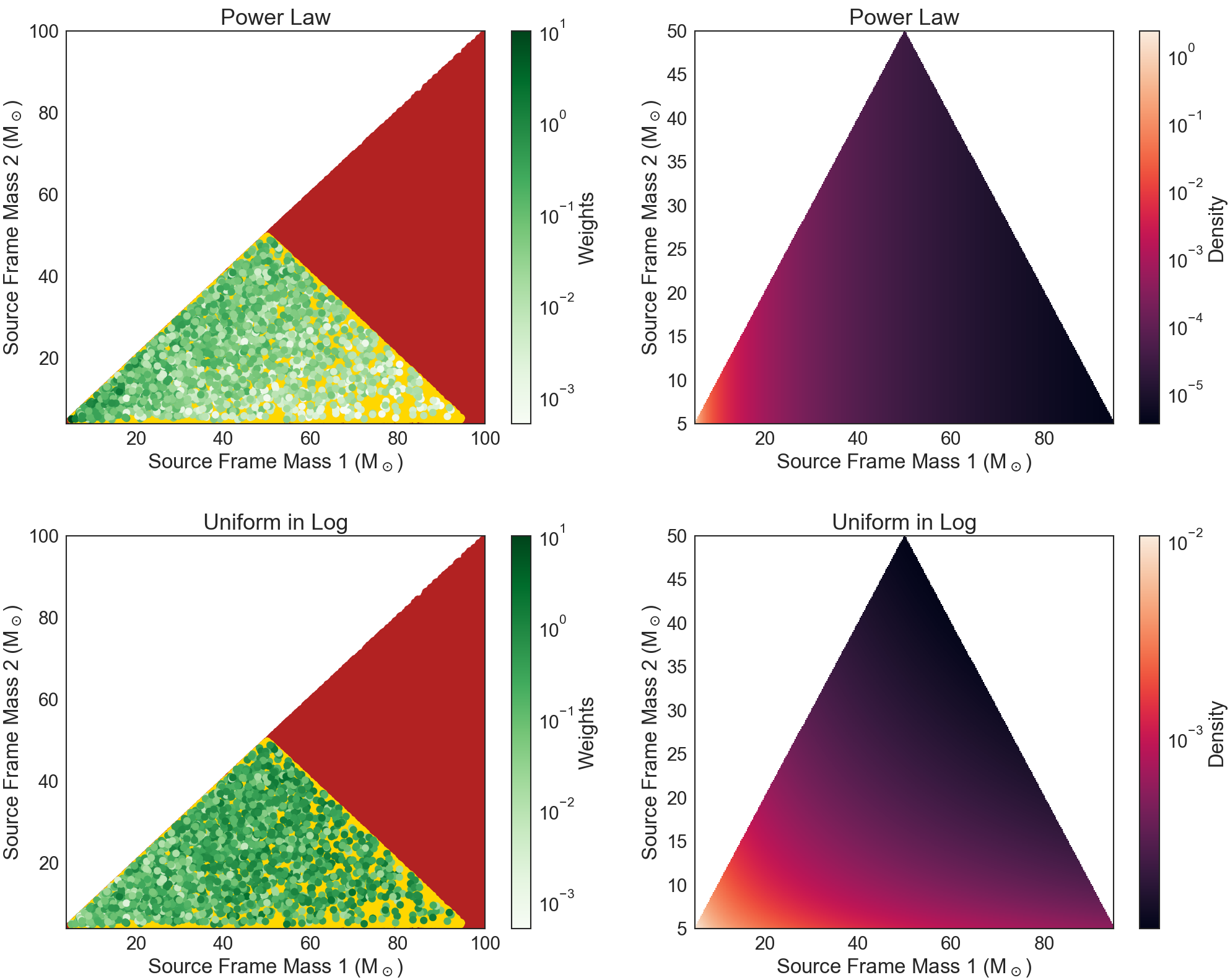}
\caption{The red dots in the figure plot the source frame masses of all the injections. Yellow dots are the source frame masses for the injections which have a non-zero weight. The green dots are masses of the recovered injections which have a color depth coding based on the value of the weights. In the plots the heavier of the two black holes is plotted as the first component mass. The maximum injected component mass is 180 M$_\odot$, but for the sake of clarity, the axes have been truncated. Additionally, the weights for the yellow dots are not shown. The right hand side of the figure plots the probability density of the two population models. Injections adequately cover the population models on the source frame component mass plane.}
\label{fig:scatter}
\end{figure}

Figure \ref{fig:scatter} shows that the injection sets adequately cover the two population models on the component mass plane. Figure also plots the weights on the \ac{MC} integration for the recovered injections. The injections follow a uniform distribution in chirp distance with maximum fiducial distance set to 300 Mpc. Current sensitivity of the detector network is less than 200 Mpc for 1.4 -- 1.4 M$_\odot$ \ac{BNS}. The maximum distance scales to around 4 Gpc for a GW150914 like binary. This maximum distance is much larger than the distances to which GW150914 like binary can be observed. Similar calculation for other \ac{GW} observation suggests that the maximum luminosity distance to which injections are placed is much larger than the current reach for the detectors. The affect of mass ratios and spins on this assessment is negligible.
Table \ref{table:comparison} compares the time volume product for the two models obtained using both the methods. 
\begin{table}
\begin{tabular}{ccc}
\hline
 Population & 100 $\langle VT \rangle$  & 100 $\langle VT \rangle$ \\
 Model & direct injections & scaled injections \\
\hline
 Uniform in log & 4.61 $\pm$ 0.1 & 4.54 $\pm$ 0.13  \\
\hline
 Power-law & 1.38 $\pm$ 0.03 & 1.36 $\pm$ 0.06 \\
\hline
\end{tabular}
\caption{Comparison of time volume product, in units of Gpc$^3$ - yr, estimated by using injections runs drawn from population models and by scaling injections in a generic injection set. The result correspond to 
calculation of rates corresponding to the event GW170104 \cite{GW170104}.}
\label{table:comparison}
\end{table}
Although the mean values are close, the statistical errors are larger for the case of weighted MC integration. Increasing the number of injections has a direct impact on the error. The error will reduce approximately as 
$1/\sqrt{F}$, where $F$ is the factor by which number of injections is increased. In fact, we expect when analyzed over the whole data obtained during LIGO's second observation run, statistical errors will reduce to around 
1\%. 

The results obtained are promising and offer a way to estimate sensitive volume without enormous investment of user and computational time. The method is open to include a larger parameter space (precession, eccentricity, 
tides etc.) or phenomena like redshift dependence of star formation rates, etc. Sensitive volume is an important ingredient when performing population inference and the estimation of corresponding merger rates. Analytical 
models can be used to estimate sensitive volume but they are usually not accurate. Strength of the method lies in its fast and accurate calculation of sensitive volume every time parameters defining a population model 
are changed (for example see \cite{GW150914AstroImp} and supplement material for \cite{GW170104}).

\section*{Acknowledgement}
The author would like to thank Gregory Mendel for providing very useful feedback on the manuscript, to Tom Dent for discussion on the injection distributions, to Albrecht Rudiger for multiple editorial fixes and to Stephen 
Fairhurst for multiple useful discussions on the topic. This work was supported by the STFC grant ST/1990s/1.

The authors thank to the LIGO Scientific Collaboration for access to the data and gratefully acknowledge the support of the United States National Science Foundation (NSF) for the construction and operation of the LIGO 
Laboratory and Advanced LIGO as well as the Science and Technology Facilities Council (STFC) of the United Kingdom, and the Max-Planck-Society (MPS) for support of the construction of Advanced LIGO. Additional support for 
Advanced LIGO was provided by the Australian Research Council.


\end{document}